\documentclass{article}
\usepackage{graphicx}

\begin{document}

\title{The role of pressure anisotropy on the maximum mass of cold compact stars}
\author{
{\bf S. Karmakar\thanks{karma@iucaa.ernet.in} ~and~S. Mukherjee\thanks{sailom@iucaa.ernet.in}}\\
Department of Physics, North Bengal University\\
Darjeeling 734 430, India\\
{\bf R. Sharma\thanks{206526115@ukzn.ac.za} ~and~S. D. Maharaj\thanks{maharaj@ukzn.ac.za}}\\
Astrophysics and Cosmology Research Unit,\\ School of Mathematical Sciences,\\
University of KwaZulu-Natal,\\ Private Bag X54001, Durban 4000, South Africa\\
}
\date{}
\maketitle

\begin{abstract}
We study the physical features of a class of exact solutions for cold compact
anisotropic stars. The effect of pressure anisotropy on the maximum mass
and surface redshift is analysed in the Vaidya-Tikekar model. It is shown that maximum compactness,
redshift and mass increase in the presence of anisotropic pressures; numerical values are generated
which are in agreement with observation.
\end{abstract}


\section{Introduction}
The description of very compact astrophysical objects has been a key issue in relativistic astrophysics
for the past decades. Recent observations suggest that there are many compact objects such as
X-ray pulsar Her X-1, X-ray burster 4U 1820-30, millisecond pulsar SAX J 1808.4-3658, X-ray sources
4U 1728-34, PSR 0943+10 and RX J185635-3754, whose estimated masses and radii are not compatible with
the standard neutron star models. The conjecture that quark matter might be the true ground state
of hadrons \cite{Witten, Farhi}, inspired many authors to describe such stars as strange
stars \cite{Kettner, Dey01}, quark-diquark stars   \cite{Horvath01}, hybrid stars \cite{Mahes} and
boson/boson-fermion stars \cite{Ruffini,Kaup,Colpi,Jetzer,Henriques01}.

As densities of such compact objects are normally above nuclear
matter density, theoretical studies suggest that pressures within
such stars are likely to be anisotropic, i.e., at the interior of
such stars there are two different kinds of pressures, viz., the
radial pressure and the tangential pressure \cite {Herrera}.
Different solutions of Einstein's field equations for anisotropic
fluid distribution with spheroidal geometry, with varying forms of the energy density, have been obtained by
many workers \cite{ Thomas, Patel, Maartens, Gokhroo}. So far,
the role of pressure anisotropy has been extensively studied in
the context of high redshift values and stability of compact
objects (see for example \cite{ Mak, Dev, Chaisi} and references
therein). Bowers and Liang \cite{Bowers} have pointed out that
anisotropy may also change the limiting values of the maximum
mass of compact stars. The objective of the present work is to
investigate the role of pressure anisotropy on the maximum masses
of compact objects. To this end, in \S 2, we modify a
solution obtained by Mukherjee {\em et al} \cite{Mukherjee} to
incorporate anisotropy. The class of solutions, capable of
describing cold compact stars, was obtained by using an ansatz
given by Vaidya and Tikekar \cite{Vaidya}. For physically
relevant anisotropic stars, the regularity and matching
conditions for the solutions are developed. In \S 3 we
discuss the role of anisotropy and calculate the maximum possible
masses for this class of solutions in \S 4. We conclude by
summarizing our results in \S 5.

\section{Anisotropic model}
We take the line element for a static spherically symmetric cold compact star in the standard form
\begin{equation}
ds^{2} = -e^{2\gamma (r)}dt^{2}+e^{2\mu (r)}dr^{2}+r^{2}(d\theta
^{2}+\sin ^{2}\theta d\phi ^{2}) \label{eq1}
\end{equation}
where $\gamma(r)$ and $\mu(r)$ are the two unknown metric functions. Assuming the energy momentum tensor for an anisotropic star in the most general form
\begin{equation}
T_{ij} = \mbox{diag}~(-\rho,~ p_{r},~ p_{\perp},~p_{\perp}) \label{eq2}
\end{equation}
the field equations are obtained as
\begin{eqnarray}
\rho &=& \frac{\left(1-e^{-2\mu}\right)}{r^2}+\frac{2\mu'e^{-2\mu}}{r}, \label{eq3}\\
p_{r} &=& \frac{2\gamma'e^{-2\mu}}{r}-\frac{\left(1-e^{-2\mu}\right)}{r^2}, \label{eq4}\\
\Delta e^{2\mu } &=& \gamma ^{\prime \prime }+{\gamma ^{\prime }}^{2}-\gamma ^{\prime
}\mu ^{\prime }-\frac{\gamma ^{\prime }}{r}-\frac{\mu ^{\prime
}}{r}-\frac{\left( 1-e^{2\mu }\right) }{r^{2}}, \label{eq5}
\end{eqnarray}
where we have set $p_{\perp}-p_{r} = \Delta$. In (\ref{eq3})-(\ref{eq5}), $\rho$ is the energy density, $p_{r}$ is the radial pressure, $p_{\perp}$ is the tangential pressure and $\Delta$ is the measure of pressure anisotropy in this model. To solve this system we use the ansatz \cite{Vaidya}
\begin{equation}
e^{2\mu }=\frac{1+\lambda r^2/R^2}{1-r^2/R^2},~ \Psi = e^{\gamma(r)}, ~ x^{2} = 1-\frac{r^2}{R^2}. \label{eq6}
\end{equation}
Then (\ref{eq5}) takes the form
\begin{equation}
(1+\lambda-\lambda x^2)\Psi_{xx} + \lambda x \Psi_{x} + \lambda
(\lambda + 1)\Psi - \frac{\Delta R^2 (1+\lambda - \lambda
x^2)^2}{(1-x^2)} \Psi = 0. \label{eq7}
\end{equation}
To solve (\ref{eq7}) we assume that the form of the anisotropic
parameter $\Delta$ is
$$\Delta = \frac{\alpha\lambda^2(1-x^2)}{R^2(1+\lambda-\lambda x^2)^2},$$
and if we make a further transformation $z = \sqrt{\lambda/(\lambda +1)} x$, (\ref{eq7}) becomes
\begin{equation}
(1-z^2)\Psi_{zz} + z\Psi_{z} + (\Lambda + 1)\Psi  = 0 \label{eq8}
\end{equation}
where $\alpha =1-\Lambda/\lambda$ is a constant. This has the general solution (see \cite{Mukherjee} for details)
\begin{equation}
e^{\gamma }=A\bigg[{\frac{\cos [(\beta+1)\zeta +\delta
]}{\beta+1}}-{\frac{ \cos [(\beta-1)\zeta
+\delta]}{\beta-1}}\bigg] \label{eq9}
\end{equation}
where, $\beta = \sqrt{\Lambda + 2}$, $\zeta =\cos ^{-1}z$, and $A$
and $\delta$ are constants which can be determined from the
boundary conditions. The physical parameters in this model are then obtained as
\begin{eqnarray}
\rho &=& {1 \over R^2 (1-z^2)} \bigg[ 1 + {2 \over (\lambda + 1) (1
- z^2)} \bigg] \label{eq10},\\
p_{r} &=& - {1 \over R^2 (1-z^2)} \bigg[ 1 + {2z \Psi_{z}\over
(\lambda + 1) \Psi} \bigg] \label{eq11},\\
p_{\perp} &=& p_{r} + \Delta \label{eq12},\\
\Delta &=& {\alpha\lambda \over R^2} \bigg[ {(\lambda + 1)(1 - z^2) - 1 \over (\lambda + 1)^2(1 - z^2)^2} \bigg], \label{eq13}
\end{eqnarray}
which together with (\ref{eq6}) and (\ref{eq9}) comprise an exact solution to the Einstein field equations. Note that
\begin{equation}
M(b) = \frac{(1+\lambda)b^3}{2R^2(1+\lambda\frac{b^2}{R^2})} \label{eq14}
\end{equation}
is the total mass  of a star of radius $b$.

We impose the following conditions in our model:
\begin{itemize}
\item At the boundary of the star the interior solution
should be matched with the Schwarzschild exterior solution, i.e.,
\begin{equation}
e^{2\gamma(r=b)} = e^{-2\mu(r=b)} = \left(1 - \frac{2M}{b}\right). \label{eq15}
\end{equation}
\item The radial pressure $p_{r}$ should vanish at the boundary of
the star which gives
\begin{equation}
\frac{\Psi_{z}(z_{b})}{\Psi(z_{b})} = - \frac{(1+\lambda)}{2z_{b}} \label{eq16}
\end{equation}
where $z_{b}^2 =  (\lambda/(\lambda +1))(1 - b^2/R^2)$. From (\ref{eq9}) we have
\begin{equation}
\frac{\psi{_z}}{\psi} = \frac{(\beta^2-1)}{\sqrt{(1-z^2)}}\left[\frac{\sin[(\beta-1)\zeta
+\delta]-\sin[(\beta+1)\zeta
+\delta]}{(\beta+1)\cos[(\beta-1)\zeta
+\delta]-(\beta-1)\cos[(\beta+1)\zeta +\delta]}\right]. \label{eq17}
\end{equation}
Combining (\ref{eq16}) and (\ref{eq17}) we obtain
\begin{equation}
\tan\delta = \frac{\tau \cot\zeta{_b} - \tan(\beta\zeta{_b})}{1+\tau
\cot\zeta{_b} \tan(\beta\zeta{_b})} \label{eq18}
\end{equation}
where $ \tau =\frac{\lambda(1 - 2\alpha)+ 1}{\beta(1 + \lambda)}$ and $\zeta{_b} = \cos ^{-1}z{_b}$.
\item $p_{r} \geq 0$ inside the star gives
\begin{equation}
\frac{\Psi_{z}}{\Psi} \leq - \frac{(1+\lambda)}{2z}. \label{eq19}
\end{equation}
\item Using (\ref{eq10})-(\ref{eq12}) we get
\begin{eqnarray}
\frac{dp_{r}}{d\rho} &=& \frac{z(1-z^2)^2(\Psi_{z}/\Psi)^2 -
(1-z^2)\Psi_{z}/\Psi) - \alpha\lambda z (1 - z^2)}{z(1-z^2)(1+\lambda) + 4z}\\ \label{eq20}
\frac{d p_{\perp}}{d\rho} &=& \frac{d p_{r}}{d\rho} + {\alpha\lambda
\over (1 + \lambda)} \bigg[ {(\lambda + 1)(1 - z^2) - 2 \over (\lambda + 1)(1 - z^2) + 4} \bigg]. \label{eq21}
\end{eqnarray}
We choose the parameters so that the causality conditions are not violated, i.e.,  $\frac{dp_r}{d\rho},~  \frac{dp_{\perp}}{d\rho} \leq 1$ in this model.
\end{itemize}
The above conditions are imposed for a physically reasonable model.

\section{Physical applications}
It was shown earlier by Sharma {\em et al} \cite{RS01} that the Vaidya-Tikekar model provides a simple method of studying systematically the maximum mass problem of compact isotropic ($\alpha=0$) stars. To see the effect of anisotropy ($\alpha \neq 0$) in this model, we may adopt the following methods.

\begin{figure}
\includegraphics{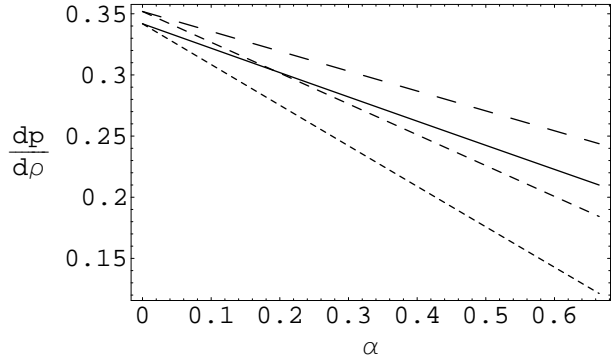}
\caption{Variations of $\frac{dp}{d\rho}$ at the boundary and at the centre of an anisotropic star against $\alpha$. We took
$\lambda = 2$ and  $u_{i}=0.1686$. The solid line is for $(\frac{dp_{r}}{d\rho})_{r=0}$, the
 dotted line is for $(\frac{dp_{\perp}}{d\rho})_{r=0}$, the long dashed line is for $(\frac{dp_{r}}{d\rho})_{r=b}$, and the dashed line is for $(\frac{dp_{\perp}}{d\rho})_{r=b}$.}
\label{fg1}
\end{figure}

\begin{figure}
\includegraphics{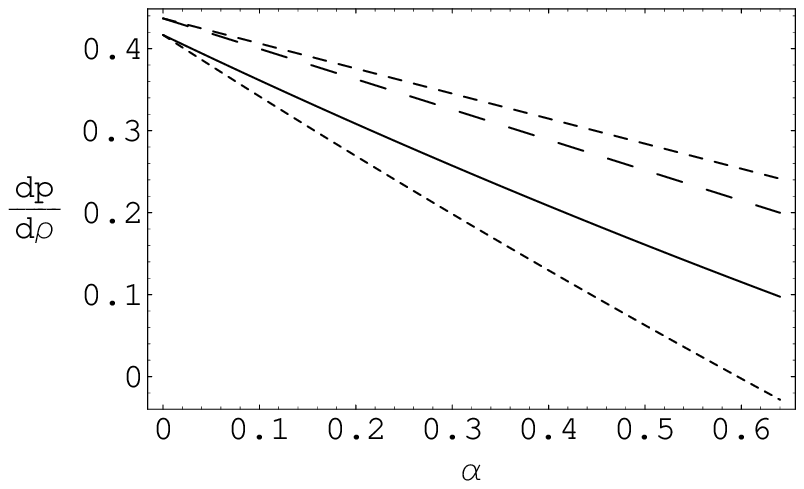}
\caption{Variations of $\frac{dp}{d\rho}$ at the boundary and at the centre of an anisotropic star against $\alpha$. We took $\lambda = 53.34$ and  $u_{i}=0.2994$. The solid line is for $(\frac{dp_{r}}{d\rho})_{r=0}$, the dotted line is for $(\frac{dp_{\perp}}{d\rho})_{r=0}$, the long dashed line is for $(\frac{dp_{r}}{d\rho})_{r=b}$, and the dashed line is for $(\frac{dp_{\perp}}{d\rho})_{r=b}$.}
\label{fg2}
\end{figure}

We may choose the isotropic compactness $u_{i}= (M/b)_{iso}$ and $\lambda$ as input parameters and using (\ref{eq14}) calculate $y = b^2/R^2$. For a given central or surface density, (\ref{eq10}) can be used to calculate the value of $R$ which then determines the radius $b = R\sqrt y$ or mass $M$ (from (\ref{eq14})). The parameter $\delta$ is fixed by choosing a specific value of $\alpha$. Since mass and radius are fixed, this method is not suitable to analyze the role of anisotropy on the maximum mass problem. However, (\ref{eq20}) and (\ref{eq21}) can be utilized to show that for stars with the same masses and radii may have different anisotropic compositions if the equations of state are modified accordingly. In \S 3.1, we consider two such examples to show how the composition may change in the presence of anisotropy. Note that the variations of the slopes of $\frac{d p_{r}}{d\rho}$ and $\frac{dp_{\perp}}{d\rho}$ may correspond to different material compositions within the star. These variations are shown in figure \ref{fg1} and figure \ref{fg2}.

To see the effect of anisotropy on the compactness, we adopt a different approach. Note that (\ref{eq18})
is a relation between $y$ and $\delta$ which we will utilize to calculate $\delta$ for given values of $\alpha$ and $\lambda$. For given values of $\lambda$ and $u_{i}$, we first calculate $y$ using (\ref{eq14}). Substituting these values in (\ref{eq18}), we determine $\delta$ for the isotropic case ($\alpha = 0$). Once $\delta$ is determined, we use this value to calculate $y_{ani}$ for different $\alpha$ values.
We then use the relation $$u_{ani} = \frac{(1 + \lambda)y_{ani}}{2(1 + \lambda y_{ani})}$$ to see the effect of anisotropy on the compactness of a star.

\subsection{Numerical results}
Following the method discussed above, we have obtained  numerical results showing the effect of anisotropy on some physically relevant parameters. Two different cases have been studied.\\
{\bf Case I:} We use our earlier data for the pulsar Her X-1 \cite{RS02} and choose $\lambda = 2$, $M = 0.88~ M_{\odot}$, $b = 7.7$~km so that $u_{i} = 0.1686$ and calculate the compactness for different anisotropic parameters.\\
{\bf Case II:} We consider the millisecond pulsar SAX J 1808.4-3658 and use the results obtained in an earlier work \cite{RS03} and choose $\lambda = 53.34$, $M = 1.435~M_{\odot}$, $b = 7.07$~km so that $u_{i} = 0.2994$.

\begin{figure}
\includegraphics{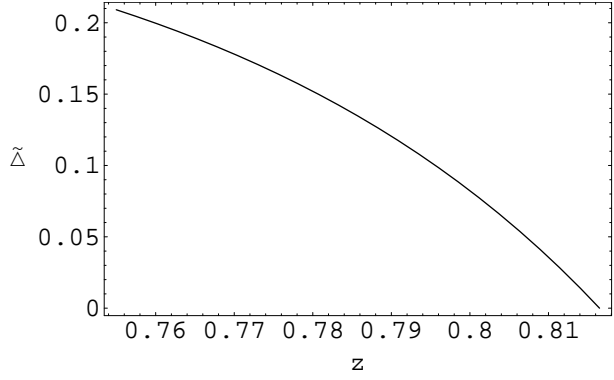}
\caption{Variation of anisotropy factor ($\tilde\Delta = R^2 \Delta$) against radial parameter $z$ for $u_{i}=0.1686$, $\lambda=2$ and $\alpha=0.6$.}
\label{fg3}
\end{figure}

We note that the compactness decreases with increasing anisotropy which is in agreement with earlier results obtained in \cite{Mak}. The results are shown in table 1. The behaviour of the anisotropy factor ($\tilde\Delta = R^2 \Delta$) in the stellar interior is shown in figure (\ref{fg3}).

\begin{table}
\begin{center}
\begin{tabular}{|l|l|l|l|l|l|l|r|}  \hline
\multicolumn{4}{|c|}{$\lambda = 2$, $b = 7.7$~km} &
\multicolumn{4}{|c|}{$\lambda = 53.34$, $b = 7.07$~km} \\ \hline
$\alpha$ &  $y_{ani}$ & $ u_{ani}$ & $M_{ani}$ ($M_{\odot}$) & $\alpha$ & $y_{ani}$
& $ u_{ani}$ & $M_{ani}$ ($M_{\odot}$) \\ \hline
0 & 0.1450 & 0.1686& 0.88&  0&  0.0267& 0.2994& 1.435  \\ \hline
0.2& 0.1281& 0.1530& 0.80& .05& 0.0269& 0.3002& 1.439 \\ \hline
0.4& 0.0953& 0.1201& 0.63& 0.1& 0.0270& 0.3006& 1.441 \\ \hline
0.5& 0.0690& 0.0909& 0.47& 0.2& 0.0268& 0.2997& 1.437 \\ \hline
0.6& 0.0322& 0.0454& 0.24& 0.4& 0.0239& 0.2858& 1.368 \\ \hline
0.65&0.0082& 0.0121& 0.06& 0.6& 0.0090& 0.1760& 0.844 \\ \hline
\end{tabular}
\caption{Compactness and mass calculated for different anisotropic parameters for two different cases discussed in
\S 3.1. }
\end{center}
\end{table}

\begin{table}
\begin{center}
\begin{tabular}{|l|l|l|l|l|l|r|}  \hline
$\alpha$   &  $y_{max}$ & $ u_{max}$ & $Z_s|_{max}$ & $b = 10$~km & \multicolumn{2}{|c|}{$\rho_{b}= 2~\rho_{nucl}$}\\ \hline
 &  &  &  &  $M_{max}$ ($M_{\odot}$) & $M_{max}$ ($M_{\odot}$) & $b_{max}$ (km) \\ \hline
0 &  0.0252 & 0.3615 &  0.9003 & 2.45 & 2.60 & 10.62 \\  \hline
0.2 &  0.0285 & 0.3738 & 0.9910 & 2.53 & 2.69 & 10.62 \\  \hline
0.4 &  0.0322 & 0.3852 & 1.0872 & 2.61 & 2.77 & 10.62  \\  \hline
0.6 &  0.0361 & 0.3955 & 1.1879 & 2.68 & 2.84 & 10.62  \\  \hline
0.7 &  0.0383 & 0.4003 & 1.2398 & 2.71 & 2.88 & 10.61  \\ \hline
0.8 &  0.0404 & 0.4048 & 1.2927 & 2.74 & 2.91 & 10.60  \\  \hline
0.9 &  0.0427 & 0.4092 & 1.3463 & 2.77 & 2.93 & 10.59 \\  \hline
1.0 &  0.0450 & 0.4132 & 1.3998 & 2.80 & 2.96  & 10.58 \\ \hline
\end{tabular}
\caption{Maximum compactness ($u_{max}$), maximum surface redshift ($Z_s|_{max}$) and maximum mass
($M_{max}$) for different anisotropic parameters . We have considered a star of radius $10$~km and surface density equal to twice $\rho_{nucl}$, where, $\rho_{nucl} = 2.7\times 10^{14}~$gm/cm$^{3}$.}
\end{center}
\end{table}

\section{Maximum mass and Surface redshift}
In an earlier work \cite{RS01}, we calculated the maximum mass for a class of isotropic stars. Here we follow the same technique to calculate the maximum mass in the presence of pressure anisotropy.
\begin{itemize}
\item We assume that $\frac{d p{_r}}{d\rho} \leq 1$ and the value is maximum at the centre. This gives
\begin{equation}
\frac{\psi_{z}}{\psi}|_{zo} \geq \frac{(1+\lambda)}{2\sqrt{\lambda}}
\left[\sqrt{\lambda+1} - \sqrt{21\lambda+1 + \frac{4 \alpha \lambda^2}{\lambda + 1}}\right] \label{eq22}.
\end{equation}
Combining (\ref{eq17}) and (\ref{eq22}), we determine the
limiting value of $\delta$ for different $\alpha$ values for a chosen value of $\lambda$.
\item Corresponding to the limiting value of $\delta$, (\ref{eq18}) can be used to
calculate the maximum value of $ y = b^2/R^2$.
\item From (\ref{eq14}) the compactness of a star in this model is given by
\begin{equation}
u = \frac{M(b)}{b} = \frac{(1+\lambda)}{2(\lambda + \frac{1}{y})}.
\end{equation}
Clearly, the maximum value of $y$ corresponds to the maximum compactness of the configuration. The maximum surface redshift ($Z_s|_{max}$) corresponding to this value can also be obtained using the following equation
\begin{equation}
Z_s|_{max}=\left(1-2u_{ani}\right)^{-1/2} -1.
\end{equation}
\end{itemize}
Once the value of maximum compactness is obtained, the maximum mass of anisotropic star can be calculated for a given radius or surface density. In \cite{RS01} we observed that for a particular choice ($\lambda = 100$), the maximum compactness for an isotropic star is $0.3615$. Keeping the same value of $\lambda$ if we go on increasing $\alpha$ we see that the maximum compactness, maximum surface redshift and maximum mass all increase with anisotropy. The results
are shown in table 2. For $\alpha$ close to unity (the maximum value of $\alpha$ in the present model is $1$) these values are almost $0.4, 1.4$ and $2.8 M_{\odot}$, respectively, for a star of radius $10$~km. These values are similar to the results obtained in \cite{Hernandez}. The maximum surface redshift obtained by Bondi
\cite{Bondi} was $1.352$ which is also very close to our values. The maximum mass for an isotropic star of radius $10$~km was $2.45M_{\odot}$ \cite{RS01}, which increases to $2.8M_{\odot}$ in the presence of anisotropy.

\section{Discussion}
We briefly point out the behaviour of the dynamical variables in
this class of models. It is clear that the energy density $\rho$ and
the radial pressure $p_r$ are decreasing functions from the centre
to the boundary of the star. This is also true for the anisotropic
stellar models of Chaisi and Maharaj \cite{Chaisi} and Sharma {\em
et al} \cite{RS03} who have studied the same spacetime geometry. The
tangential pressure $p_{\perp}$ has a more complicated behaviour
because it is related to the anisotropy factor via $p_{\perp} = p_r
+\Delta$; in addition $p_{\perp}$ depends on the Gegenbaur function
and the new variable $z$ rather than the original radial coordinate
$r$. To illustrate the behaviour of $p_{\perp}$ we have generated a
plot in Figure 4. It is clear that the tangential pressure is an
increasing function as we approach the centre. This is physically
acceptable since the conservation of angular momentum during the
quasi-equilibrium contraction of a massive body should lead to high
values of $p_{\perp}$ in the central regions of the star.
\begin{figure}
\includegraphics{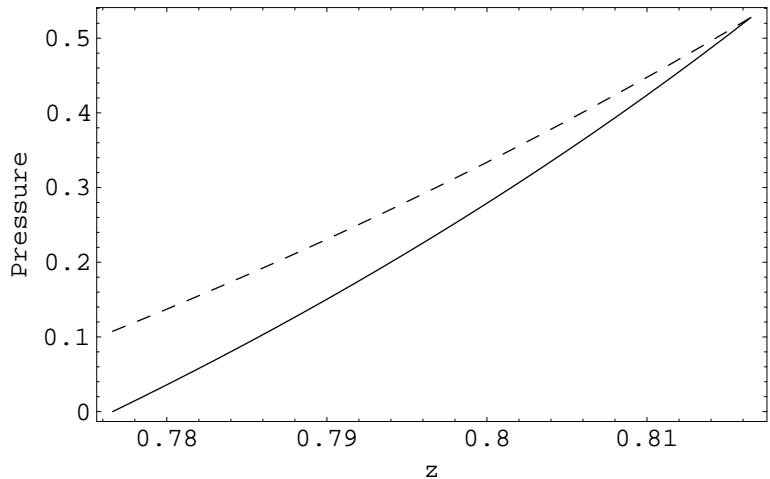}
\caption{Radial pressure (solid curve) and tangential
pressure (dashed curve) plotted against the radial parameter
$z$ for $\alpha=0.4$ and $\lambda =2$.}
\label{fg4}
\end{figure}

We have extended a class of solutions describing cold compact stars to incorporate anisotropy. The solutions were then used to see the effect of anisotropy on the maximum possible mass and surface redshift parameters of cold compact stars. A comparative study of our results with earlier results are given in table 3. The anisotropy in the present
model vanishes at the centre and reaches the maximum value at the surface of the star as shown in the figure 3.  Unlike some earlier works \cite{Dev, Chaisi}, this model has an isotropic counterpart ($\alpha = 0$) which helps to compare anisotropic stars with their isotropic counterparts. In this model we assumed $p_{\perp} >p_{r}$ and have shown that the upper bound on the maximum mass increases in the presence of anisotropy. To conclude, our model provides a simple method to fix the upper bound on the maximum possible masses for the class of compact anisotropic stars described by the Vaidya-Tikekar model.

\begin{table}
\begin{center}
\begin{tabular}{|l|l|r|}  \hline
$References$   & $ 2u_{max}$ & $Zs_{max}$ \\ \hline
Guven and Murchadha \cite{Guven} & 0.974 &  5.211   \\ \hline
Ivanov \cite{Ivanov}             & 0.957 & 3.842     \\  \hline
Bondi \cite{Bondi}                & 0.819 & 1.352     \\ \hline
Hern$\acute{a}$ndez and N$\acute{u}$$\widetilde{n}$ez
\cite{Hernandez}                   & 0.800 & 1.200       \\ \hline
Present Work                    & 0.826 & 1.400   \\ \hline
\end{tabular}
\caption{Maximum compactness ($u_{max}$) and maximum surface
redshift ($Z_s|_{max}$) of anisotropic stars in different models.}
\end{center}
\end{table}

\section*{Acknowledgment}
RS acknowledges the financial support (grant no. SFP2005070600007) from the National Research Foundation (NRF), South Africa. SDM acknowledges that this work is based upon research supported by the South African Research Chair Initiative of the Department of Science and Technology and the National Research Foundation.

\end{document}